%% file: root.tex
\documentclass[letterpaper, 10 pt, conference]{ieeeconf}  

\IEEEoverridecommandlockouts                              

\usepackage{graphics} 
\usepackage{epsfig} 
\usepackage{times} 
\usepackage{amsmath} 
\usepackage{amssymb}  

\usepackage[style=ieee]{biblatex} 
\usepackage{color} 
\usepackage{xcolor}
\usepackage{hyperref}
\usepackage{lipsum}
\usepackage{tikz}
\usetikzlibrary{arrows.meta, positioning}
\usepackage{booktabs}
\usepackage{comment}

\title{\LARGE \bf
\mods{Towards} Fair and Efficient allocation of \\ Mobility-on-Demand resources through a Karma Economy
}

\author{Matteo Cederle, Saverio Bolognani, and Gian Antonio Susto
\thanks{Matteo Cederle and Gian Antonio Susto are with the Department of Information Engineering, University of Padova, Italy.
{Emails: matteo.cederle@phd.unipd.it, gianantonio.susto@unipd.it}}%
\thanks{Saverio Bolognani is with the Automatic Control Laboratory, ETH Zurich, Switzerland. {Email: bsaverio@ethz.ch}}%
\thanks{This study was partially carried out within the Italian National Center for Sustainable Mobility (MOST) and received funding from NextGenerationEU (Italian NRRP – CN00000023 - D.D. 1033 17/06/2022 - CUP C93C22002750006), from the NCCR Automation, grant agreement 51NF40$\_$225155 of the Swiss National Science Foundation, and from the Ing. Aldo Gini Foundation.}
}

\newcommand{\mods}[1]{\textcolor{black}{#1}} 

\newtheorem{definition}{\textbf{Definition}}
\newtheorem{assumption}{\textbf{Assumption}}
\newtheorem{theorem}{\textbf{Theorem}}

\begin{document}

\maketitle
\thispagestyle{empty}
\pagestyle{empty}

\begin{abstract}
Mobility-on-demand systems like ride-hailing have transformed urban  transportation, but they have also exacerbated socio-economic inequalities in access to these services, also due to surge pricing strategies. Although several fairness-aware frameworks have been proposed in smart mobility, they often overlook the temporal and situational variability of user urgency that shapes real-world transportation demands. This paper introduces a non-monetary, Karma-based mechanism that models endogenous urgency, allowing user time-sensitivity to evolve in response to system conditions as well as external factors. We develop a theoretical framework maintaining the efficiency and fairness guarantees of classical Karma economies, while accommodating this realistic user behavior modeling. 
\mods{Applied to a simplified simulated mobility-on-demand scenario, we provide a proof-of-concept illustration of the proposed framework, showing that it exhibits promising behavior in terms of system efficiency and equitable resource allocation, while acknowledging that a full treatment of realistic MoD complexity remains an important direction for future work.}

\emph{Index Terms —} Algorithmic fairness, Game~theory, Karma~economies, Mobility on Demand, Ride-hailing.
\end{abstract}

\section{INTRODUCTION}
\label{sec:intro}
\input{1-intro}
\section{PROBLEM FORMULATION}
\label{sec:problem}
\input{2-problemformulation}
\section{AUGMENTED KARMA ECONOMIES FOR MOBILITY ON DEMAND}
\label{sec:contr}
\input{3-karma}
\section{NUMERICAL ANALYSIS}
\label{sec:exp}
\input{4-experiments}
\section{CONCLUSIONS}
\label{sec:conc}
\input{5-conclusions}
\addtolength{\textheight}{-12cm}   




\printbibliography

\end{document}

%% file: 1-intro.tex
The proliferation of Mobility-on-Demand (MoD) systems, particularly ride-hailing platforms, has fundamentally transformed urban transportation landscapes over the past decade~\cite{shaheen2020mobility, zardini2022analysis}. These systems have demonstrated remarkable utility in providing flexible and accessible transportation solutions that adapt to real-time demand patterns and user preferences \cite{liyanage2019flexible, atasoy2015concept}. However, despite their widespread adoption and operational efficiency, ride-hailing services have increasingly exposed significant socio-economic disparities in urban mobility access \cite{dholakia2015everyone, shontell2014uber}. Lower-income communities often face higher waiting times, reduced service availability, and surge pricing effects that disproportionately impact their mobility options, thereby exacerbating existing transportation inequities.~\cite{guan2024shared}

In response to these challenges, the research community has begun developing fairness-oriented approaches within the smart mobility domain. Recent work has explored algorithmic mechanisms to ensure more equitable service distribution across different demographic groups \cite{cederle2025fairness, cederle2025bfairness}, investigated demand prediction strategies that mitigate discriminatory effects \cite{yan2020fairness, zheng2023fairness}, and proposed optimization frameworks that explicitly incorporate fairness constraints in resource allocation \cite{salazar2024accessibility}. While these contributions represent important steps toward more equitable mobility systems, substantial work is still needed to develop fair and applicable strategies in this domain.

Contextually, Karma economies have emerged as a compelling theoretical paradigm that offers rigorous mathematical guarantees for addressing scarce resource allocation problems in both an efficient and fair manner \cite{censi2019today, elokda2024self}\mods{, exploiting recent results in the Dynamic Population Games (DPGs) literature~\cite{elokda2024dynamic}}. This framework is particularly relevant to mobility contexts, where the fundamental challenge lies in optimally matching transportation supply with user demand, while ensuring \mods{fairness for} diverse user populations\mods{, defined as the maximization of Long-run Nash welfare~\cite{elokda2025vision}, a criterion that does not require interpersonal comparability of utilities (see \cite{elokda2024self}, Section~3.4)}. However, the current Karma economy framework has a key limitation: it treats user urgency as entirely exogenous, meaning that the time-sensitivity of transportation requests depends solely on external factors and not on the evolving state of the system. This assumption fails to capture the reality of MoD scenarios, where user urgency is inherently endogenous—influenced by factors such as real-time traffic conditions, available alternatives, and dynamic pricing mechanisms. \mods{We note that a fully realistic MoD model would additionally require capturing geographical mismatches between users and vehicles, rebalancing trips, and spatially heterogeneous waiting times, which would necessitate richer environment state representations and more complex reward functions. Addressing these aspects is beyond the scope of this paper; rather, our goal is to take a first step toward integrating endogenous urgency into the Karma framework, as a necessary theoretical building block for future, more comprehensive MoD applications.}

This paper proposes a theoretical extension of Karma economies that incorporates endogenous urgency processes for users, thereby creating a more realistic and practically applicable framework for \mods{simplified} MoD scenarios such as ride-hailing. By allowing user urgency to evolve dynamically based on system state and individual circumstances, our approach better reflects \mods{some of the} complex decision-making processes that characterize real transportation networks, while maintaining the theoretical rigor and fairness guarantees of the original Karma framework.

The primary contributions of this paper are threefold:

\begin{itemize}
    \item We propose a simplified fairness-oriented ride-hailing allocation problem, along with the key features that distinguish it from traditional mechanism design approaches in transportation systems.
    \item We extend the Karma economy framework to account for scenarios in which user urgency is endogenous, evolving in response to the system rather than being dictated solely by external factors, while preserving and extending the theoretical guarantees of efficiency and fairness that characterize the original framework.
    \item We demonstrate the applicability of this enhanced theoretical framework through its implementation in a simulated scenario, \mods{illustrating how the augmented Karma economy behaves under endogenous urgency and providing a proof-of-concept for its potential use in fair ride-hailing allocation problems.}
\end{itemize}

The remainder of this manuscript unfolds as follows. Section~\ref{sec:problem} defines the system model and the specific characteristics of the problem, while Section~\ref{sec:contr} introduces the game-theoretic approach for Karma-based ride-hailing systems, considering endogenous urgency processes for users. To support the theoretical findings, Section~\ref{sec:exp} reports on a case
study. Lastly,
conclusions and future outlooks are reported in Section~\ref{sec:conc}.
\subsection{Notation}
Let $D$ be a discrete set and $C$ be a continuous set. Let $a,d \in D$ and $c \in C$. For a function $f : D \times C \rightarrow~\mathbb{R}$, we distinguish the discrete and continuous arguments through the notation $f[d](c)$. Moreover, $g[a|d](c)$ denotes the conditional probability of $a$ given $d$ and $c$. Specifically, $g[d^+|d](c)$ denotes one-step transition probabilities for $d$. We denote by $\Delta(D):=\left\{\left. p \in \mathbb{R}_+^{\lvert D \rvert} \right\rvert \sum_{d \in D} p[d] = 1 \right\}$ the set of probability distributions over the elements of $D$. For a probability distribution $p \in \Delta(D)$, $p[d]$ denotes the probability of element $d$.

%% file: 2-problemformulation.tex
In this section, we formalize our simplified ride-hailing allocation problem. We begin by establishing the fundamental structure of the problem, then articulate the key characteristics that distinguish this formulation from traditional approaches in transportation systems.

\subsection{System Model}
\label{subsec:sysmod}
Consider a MoD system operating over discrete time instants $t \in \mathbb{N}$, with a finite but large set of users $\mathcal{N} =~\{1, \ldots, N\}$, and a fleet of $M\ll N$ vehicles. The system must allocate available vehicles to pending requests, while managing at the same time the inherent scarcity of transportation resources.

We consider a deliberately simple interaction model, at every discrete time $t$ two users compete for the closest ride-hailing driver.
Our objective is to design a mechanism that allocates the resource -- i.e., grants the trip request -- to one of the two users.
The users are equipped with a private time-varying attribute $x=[u,c]$, where $u\in\mathcal{U}=~\{u_1,...,u_{n_u}\}$\footnote{Note that $u_i\in\mathbb{N}$ and $u_i<u_{i+1}$ for $i=1,\dots n_u-1$.} is the urgency state, representing a private valuation for the resource at every time instant, and $c$ denotes a tradable currency or token—the specific form of which (money, karma points, credits, etc.) depends on the allocation mechanism under consideration.
The outcome of the interaction is determined by some \textit{resource allocation} and \textit{payment rules}, which are also designed depending on the specific method.

\subsection{Problem-specific characteristics}
The problem we address is characterized by some peculiar features that, taken together, create a fundamentally different allocation challenge compared to standard mechanism design problems. We start from the fact that users possess private urgency information, not directly observable by the service providers. A user traveling to a medical appointment or to a job interview experiences fundamentally different urgency levels than one making a leisure trip, even if the trips appear identical in terms of observable characteristics, such as distance and time of the day.
This private information creates a mechanism design challenge: users must be incentivized to truthfully report their urgency, yet the service provider cannot independently verify these claims.

The aforementioned peculiarity of our problem is strictly related to the fact that the urgency processes of the users are endogenous to the system dynamics, meaning that they evolve as a function of strategic interactions and waiting time. Mathematically speaking, we can model the urgency at consecutive resource competition instances for a specific user as following an irreducible Markov chain process $\Phi[u^+|u,o]$, where $o\in\mathcal{O}=\{0,1\}$ denotes the resource competition outcome for the user, i.e., $o=1$ means that it is yielding, while $o=0$ means that it is selected to access the resource.

Finally, the societal challenges mentioned in Section~\ref{sec:intro} and the multi-stakeholder nature of modern urban mobility systems require the development of fairness-oriented approaches within the urban mobility domain. While traditional ride-hailing platforms employ monetary instruments to balance supply and demand, these mechanisms have demonstrated significant regressive effects that contradict the broader social objectives of public transportation policy \cite{dholakia2015everyone}.
Municipal authorities and public transit agencies increasingly recognize that mobility access represents a fundamental dimension of urban equity, directly impacting employment opportunities, healthcare access, education, and social participation. When ride-hailing services effectively complement or substitute for public transit---particularly in under-served areas or during off-peak hours---their allocation mechanisms carry implications that extend far beyond private market efficiency.

%% file: 3-karma.tex
This section presents a solution for the fairness-oriented ride-hailing problem outlined in Section~\ref{sec:problem}, by adapting the traditional Karma economy framework to capture the dynamic, system-dependent urgency of users in MoD settings.

\subsection{A novel Karma economy for ride-hailing systems}
\label{subsec:model}
\begin{figure}[tb]
\begin{tikzpicture}[
    node distance=1.7cm,
    every node/.style={circle, draw, minimum size=1cm, font=\small, align=center},
    forward/.style={-{Latex[length=1.5mm, width=1mm]}, thin},
    backward/.style={-{Latex[length=1.5mm, width=1mm]}, bend right=40, thin},
    self/.style={-{Latex[length=1.5mm, width=1mm]}, thin, looseness=6, out=120, in=60}
]

\node (A) {$u_1$};
\node (B) [right of=A] {$u_2$};
\node (C) [right of=B] {$u_3$};
\node (D) [right of=C] {$u_4$};
\node (E) [right of=D] {$u_5$};

\draw[forward] (A) -- (B);
\draw[forward] (B) -- (C);
\draw[forward] (C) -- (D);
\draw[forward] (D) -- (E);

\draw[backward, dashed] (B) to (A);
\draw[backward, dashed] (C) to (A);
\draw[backward, dashed] (D) to (A);
\draw[backward, dashed] (E) to (A);

\draw[self, dashed] (A) to[out=120, in=60, looseness=6] (A);
\draw[self] (E) to[out=120, in=60, looseness=6] (E);

\end{tikzpicture}
\caption{Graphical representation of $\Phi[u^+|u,o]$. The solid arrows represent the high probability transitions when $o=1$, while the dashed arrows represent the high probability transitions when $o=0$.
}
\label{fig:graph1}
\end{figure}
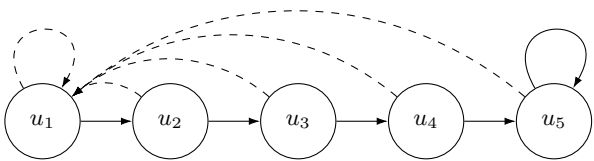

Karma economies are modeled on the class of \textit{discrete-time, finite-state-and-action, stationary mean-field games}, also known as Dynamic Population Games (DPGs), as introduced in \cite{elokda2024dynamic}. As usual in mean-field games, we take the perspective of an \textit{ego agent} playing against the population, from which a random opponent is drawn at each resource competition instance. The ego agent is equipped with a private time-varying state $x=[u,k]\in\mathcal{X}=\mathcal{U}\times\mathbb{N}$, where $u\in\mathcal{U}$ has been introduced in Section~\ref{subsec:sysmod}, and $k\in\mathbb{N}$ is the current \textit{karma} of the agent, defined as an artificial currency whose total value in the system is preserved at all times, as we will formalize in Section~\ref{subsec:sne}. The urgency at consecutive resource competition instances follows an exogenous process $\Phi[u^+|u]$. As already introduced in Section~\ref{sec:intro}, this simplified assumption does not capture the reality of MoD scenarios. 
To cope with this, we introduce in this work an endogenous transition process that aims to capture the urgency levels of users that request access to ride-hailing services:
\begin{alignat}{2}
\label{eq:phi}
    &\Phi[u^+|u,o]= \nonumber\\
    &\begin{cases}
        1-\varepsilon, \quad &\text{if }u=u_i, u^+=u_1, o=0, \\
        1-\varepsilon, \quad &\text{if }u=u_i, u^+=\min(u_{i+1},u_{n_u}), o=1, \\
        \frac{\varepsilon}{n_u-1}, \hspace{0.1cm} &\text{if }u=u_i, u^+\in\mathcal{U}\symbol{92}\{u_1\}, o=0, \\
        \frac{\varepsilon}{n_u-1}, \hspace{0.1cm} &\text{if }u=u_i, u^+\in\mathcal{U}\symbol{92}\{\min(u_{i+1},u_{n_u})\}, o=1,
    \end{cases}
\end{alignat}
\normalsize
where $\varepsilon\in\mathbb{R}_{]0,1[}$, $\varepsilon\simeq0$. 
This definition encodes the typical trend of ride-hailing requests: the urgency level of a user increases (up to a saturation point) if its request is not served, and it immediately decreases to the lowest level whenever its request is satisfied.  Moreover, we designed $\varepsilon$ to be slightly higher than 0 in order to keep some stochasticity in the system, encoding unexpected situations that could arise at any time, and to represent users that do not follow the aforementioned standard urgency pattern (e.g., users traveling to a medical appointment are immediately high urgent, and conversely users making a leisure trip remain in low urgency states even though their request is not rapidly satisfied). Figure \ref{fig:graph1} graphically shows an example of the aforementioned urgency transition process, where we considered $|\mathcal{U}|=5$ for ease of visualization.

Moving forward, the joint distribution of the agents' states in the population is given by $d\in\mathcal{D}:=\{d\in\mathbb{R}_+^\infty|\sum_{u,k}d[u,k]=1\}$, where $d[u,k]$ denotes the fraction of agents in state $[u,k]$. The ego agent places a bid whose maximum value is determined by its karma, i.e., $b\in\mathcal{B}^k:=\{b\in\mathbb{N}|b\leq k\}$, chosen according to its policy $\pi:\mathcal{X}\to\Delta(\mathcal{B}^k)$, which maps the agent's state to a probability distribution over the bids. The pair $(d,\pi)$ is referred to as the \textit{social state}, as it gives a macroscopic description of the distribution of the agents' states, as well as how they behave. Since the ego agent gets matched with an opponent drawn from the population, an important quantity is the distribution of other agents' bids $\nu[b'](d,\pi)=\sum_{u',k'}d[u',k']\pi[b'|u',k']$. Moreover, we denote the conditional probability distribution for the resource competition outcome $o\in\mathcal{O}$ of the ego agent by $\mathbb{P}[o|b,b']$, behaving as follows:
\begin{alignat}{2}
\label{eq:prob}
    &\mathbb{P}[o=0|b,b']=\begin{cases}
        1,\hfill \text{if } b>b', \\
        0,\hfill \text{if } b<b', \\
        0.5,\hfill \text{if } b=b'.
    \end{cases}, \nonumber\\ &\mathbb{P}[o=1|b,b']= 1 - \mathbb{P}[o=0|b,b'].
\end{alignat}
Equation (\ref{eq:prob}) lets us define the probability of the resource competition outcome for the ego agent as
\begin{equation}
    \gamma[o|b](d,\pi)=~\sum_{b'}\nu[b'](d,\pi)\mathbb{P}[o|b,b'].
\end{equation}
\mods{Finally, we introduce the immediate reward function and state transition function as follows (see \cite{elokda2024self} for more details):}
\begin{equation}
\label{eq:reward}
    \xi[u,k,b](d,\pi)=\xi[u,b](d,\pi)=-u\gamma[o=1|b](d,\pi).
\end{equation}
\begin{equation}
\begin{aligned}
\label{eq:newstatetr}
\rho[&u^+,k^+|u,k,b](d,\pi)= \\ &=\sum_o\gamma[o|b](d,\pi)\kappa[k^+|k,b,o](d,\pi)\Phi[u^+|u,o],
\end{aligned}
\end{equation}
where the karma transition function $\kappa[k^+|k,b,o](d,\pi)$ can be defined in various ways\mods{, provided that it satisfies the following assumption:}
\begin{assumption}
\label{ass2}
    \textit{Karma is preserved for all $(d,\pi)$, when taking the expectation over the entire population, i.e.,}
    \begin{equation}
        \mathbb{E}_{\substack{[u,k] \sim d \\b \sim \pi[\cdot \mid u,k]}}[k^+] = \mathbb{E}_{[u,k] \sim d}[k].
    \end{equation}
\end{assumption}

For this work, we consider the most widely used \textit{pay bid to society} scheme \cite{elokda2024self}, where each user pays its bid if it is selected, and pays nothing otherwise. Its conditional payment is thus given by:
\begin{align}
    p[b,o]=\begin{cases}
        b, \quad &\text{if }o=0, \\
        0, \quad &\text{otherwise.}
    \end{cases}
\end{align}
The average payment across the population is then given by:
\begin{equation}
    \bar{p}(d,\pi)=\sum_{u,k}d[u,k]\sum_b\pi[b|u,k]\gamma[o=0|b](d,\pi)b.
\end{equation}
This gets redistributed to all the users through the following integer-preserving redistribution rule: distribute $\lfloor \bar{p}(d,\pi)\rfloor$ to a fraction $f^{\text{low}}(d,\pi):=\lceil \bar{p}(d,\pi)\rceil - \bar{p}(d,\pi)$ of agents, randomly selected; distribute $\lceil \bar{p}(d,\pi)\rceil$ to the remaining fraction $f^{\text{high}}(d,\pi):= 1-f^{\text{low}}(d,\pi)$ of users. Finally, the karma transition function takes the following form:
\begin{multline}
    \label{eq:kappa}
    \kappa[k^+ \mid k,b,o](d,\pi)= \\
    = \begin{cases}
        f^\textup{low}(d,\pi), &\text{if } o = 0 \text{ and } k^+ = k - b + \lfloor\bar{p}(d,\pi)\rfloor, \\
        f^\textup{high}(d,\pi), &\text{if } o = 0 \text{ and } k^+ = k - b + \lceil\bar{p}(d,\pi)\rceil, \\
        f^\textup{low}(d,\pi), &\text{if } o = 1 \text{ and } k^+ = k + \lfloor\bar{p}(d,\pi)\rfloor, \\
        f^\textup{high}(d,\pi), &\text{if } o = 1 \text{ and } k^+ = k + \lceil\bar{p}(d,\pi)\rceil, \\
        0, &\text{otherwise}.
    \end{cases}
\end{multline}
\subsection{Existence of Stationary Nash Equilibrium}
\label{subsec:sne}
If we assume that the ego agent discounts its future rewards with a factor $\alpha\in[0,1)$, we can express the \textit{expected immediate reward} of the ego agent under policy $\pi$ as:
\begin{equation}
    R[u,k](d,\pi)=\sum_b\pi[b|u,k](d,\pi)\xi[u,b](d,\pi),
\end{equation}
and its \textit{state transition probabilities} as:
\begin{equation}
    P[u^+,k^+|u,k](d,\pi)=\sum_b\pi[b|u,k]\rho[u^+,k^+|u,k,b](d,\pi).
\end{equation}
The \textit{expected infinite horizon reward} is therefore defined as:
\begin{equation}
\begin{aligned}
    V[&u,k](d,\pi)=R[u,k](d,\pi)+ \\ &+\alpha\sum_{u^+,k^+}P[u^+,k^+|u,k](d,\pi)V[u^+,k^+](d,\pi),
\end{aligned}
\end{equation}
which in turn allows us to define the ego agent's \textit{single-stage deviation reward (Q-function)}:
\begin{equation}
\begin{aligned}
    Q[u,&k,b](d,\pi)=\xi[u,b](d,\pi)+ \\ &+\alpha\sum_{u^+,k^+}\rho[u^+,k^+|u,k,b](d,\pi)V[u^+,k^+](d,\pi).
\end{aligned}
\end{equation}
Consequently, the \textit{state-dependent best response} of the ego agent is:
\begin{equation}
\begin{aligned}
    B[u,k](d,\pi)
    &\in \{\sigma \in \Delta(\mathcal{B}^k) \lvert \forall \sigma' \in \Delta(\mathcal{B}^k), \; \\ &\sum\limits_b (\sigma[b] - \sigma'[b]) Q[u,k,b](d,\pi) \geq 0\}.
\end{aligned}
\end{equation}
We are now ready to define the adopted solution concept:
\begin{definition}
    \textit{A Stationary Nash Equilibrium (SNE) is a social state $(d,\pi)$ which satisfies for all $[u,k]\in\mathcal{U}\times\mathbb{N}$:}
    \begin{equation}        d[u,k]=\sum_{u^-,k^-}d[u^-,k^-]P[u,k|u^-,k^-](d,\pi),
    \end{equation}
    \begin{equation}
    \pi[\cdot|u,k]\in B[u,k](d,\pi).
    \end{equation}
\end{definition}

To prove the existence of an SNE in the augmented Karma economy for ride-hailing systems we need to make \mods{the following assumption}:

\begin{assumption}
\label{ass}
    \textit{The state transition function $\rho[u^+,k^+|u,k,b](d,\pi)$ is continuous in $(d,\pi)$.}
\end{assumption}

\mods{To verify that
Assumption \ref{ass} holds, some additional considerations are required,} because the state transition function defined in (\ref{eq:newstatetr}) is different from the one used in the Karma games literature. However, we can easily see that also Assumption \ref{ass} is satisfied, since $\Phi[u^+|u,o]$ is independent from $(d,\pi)$, while $\gamma[o|b](d,\pi)$ and $\kappa[k^+|k,b,o](d,\pi)$ are indeed continuous in the social state $(d,\pi)$.

This leads us to the following theorem:
\begin{theorem}
    \textit{Let Assumptions \ref{ass2} and \ref{ass} hold. Then, for each $\bar{k}\in\mathbb{N}$, a Stationary Nash Equilibrium $(d^*,\pi^*)$ satisfying $\sum_{u,k}d^*[u,k]k=\bar{k}$ is guaranteed to exist in the extended Karma formulation with endogenous urgency processes.} 
\end{theorem}
The proof depends on Assumption \ref{ass2} to ensure that the set of state distributions $\mathcal{D}$ is compact, which allows the application of an infinite-dimensional version of Kakutani’s fixed point theorem (see \cite{elokda2024self}). In \cite{elokda2024dynamic}, an algorithm is developed for computing SNE in Dynamic Population Games. This algorithm is based on establishing an equivalence between the SNE and a standard Nash Equilibrium in a suitably defined static population game, and it employs standard evolutionary dynamics \cite{sandholm2010population} to compute the equilibrium. In the following section, we leverage these tools to analyze the performance of our extended Karma game formulation at the SNE in a simplified ride-hailing allocation problem.

%% file: 4-experiments.tex
In this section, we evaluate our proposed method within a \mods{simplified} simulated mobility-on-demand platform for ride-hailing services. We first present the performance metrics used for assessment, followed by a description of the benchmark methods against which our approach is compared. We then present and discuss the obtained numerical results\footnote{The code for all the experiments can be found at \url{https://github.com/mcederle99/MoD-Karma}.}.

\subsection{Performance metrics and benchmarks}
\label{subsec:bench}
In order to quantitatively assess the performance of our method, we use the long-run average reward $\bar{R}$ across the whole population at the equilibrium, and the \textit{ex-post reward fairness} $\beta$
as social welfare metrics. Given the definition of the reward function in (\ref{eq:reward}), higher scores for $\bar{R}$ indicate that the method is able to effectively grant trips to the users which necessitates them the most. Moreover, in our context we define the ex-post reward fairness as follows:
\begin{equation}
    \beta = -\texttt{std}_{i\in\mathcal{N}}\cfrac{1}{T}\sum_{t=0}^{T-1}R_i^t\leq 0
\end{equation}
where $R_i^t$ denotes the reward for user $i$ at interaction $t$\mods{, and \texttt{std} stands for standard deviation}. As we can see, higher values for $\beta$ indicate higher levels of fairness of access across the population.

In this work, we compared our Karma-based ride-hailing allocation scheme against two different methods: a trivial random baseline (\texttt{RANDOM}), which simply tosses a coin to decide which user's trip request will be served at each interaction, irrespectively of the private urgency of the different users, and a resource allocation scheme that ensures that the players take turns accessing the resource, by selecting the user who has received the resource the least fraction of times in the past (\texttt{TURN}). Moreover, we also define the upper bound for the long-run average reward in the system \texttt{MAX\_EFF}, which has access to the users' private urgency and grants the trip to the user with the highest one at each interaction.
While \texttt{RANDOM} and \texttt{TURN} are trivial to define, \texttt{MAX\_EFF} requires solving an optimization problem, defined as follows:
\begin{equation}
\begin{aligned}
\label{eq:optpb}
\max_\psi \quad & \bar{R}=\sum_{u\in\mathcal{U}}\sum_{o\in\mathcal{O}}\psi[u,o]\cdot r[u,o]\\
\textrm{s.t.} \quad & \sum_o\psi[u,o]=\sum_{u^-,o}\psi[u^-,o]\cdot\Phi[u|u^-,o],\quad \forall u\in\mathcal{U}\\
  &\sum_{u,o}\psi[u,o]=1    \\
  &\psi[u,o]\geq 0,\quad \forall (u,o)\in\mathcal{U}\times\mathcal{O} \\
  &\sum_u \psi[u,0]=0.5
\end{aligned}
\end{equation}
Here, $\psi[\cdot,\cdot]\in\mathcal{U}
\times\mathcal{O}$ denotes the joint distribution over states and outcomes, and $r[\cdot,\cdot]$ is the same reward function of the Karma model, i.e.:
\begin{align}
    r[u,o]=\begin{cases}
    0 \quad &\text{if }o=0, \\
    -u \quad &\text{if }o=1.
    \end{cases}
\end{align}
The first constraint is needed to require the stationarity of the urgency distribution, the second and third one ensure that the joint urgency-outcome distribution is a valid probability distribution, and the last constraint ensures that the ride-hailing trip is granted to one of the two users paired to compete for it. It is clear that the optimization problem in (\ref{eq:optpb}) is a Linear Program, which can be solved efficiently.

\subsection{Case studies}
\label{subsec:casestudies}
For our experiments we consider a large population composed by $N=1000$ players, which discount their future rewards with $\alpha=0.98$, possess an average amount of karma equal to $\bar{k}=10$, and participate in $T=1000$ interactions.
The effectiveness of our method in providing fair and efficient scheduling of ride-hailing trips is shown through a numerical simulation. We consider the users to have five different urgency levels, i.e. $\mathcal{U}=\{1,2,4,8,16\}$, and the following transition process, which depends on the outcome~of the~game: 
\begin{equation}
\begin{aligned}
\label{eq:phi1}
    \Phi[u^+|u,o=0]=\begin{pmatrix}
        0.96 & 0.01 & 0.01 & 0.01 & 0.01 \\
        0.96 & 0.01 & 0.01 & 0.01 & 0.01 \\
        0.96 & 0.01 & 0.01 & 0.01 & 0.01 \\
        0.96 & 0.01 & 0.01 & 0.01 & 0.01 \\
        0.96 & 0.01 & 0.01 & 0.01 & 0.01 
    \end{pmatrix}, \\
    \Phi[u^+|u,o=1]=\begin{pmatrix}
        0.01 & 0.96 & 0.01 & 0.01 & 0.01 \\
        0.01 & 0.01 & 0.96 & 0.01 & 0.01 \\
        0.01 & 0.01 & 0.01 & 0.96 & 0.01 \\
        0.01 & 0.01 & 0.01 & 0.01 & 0.96 \\
        0.01 & 0.01 & 0.01 & 0.01 & 0.96 
    \end{pmatrix}. 
\end{aligned}
\end{equation}

The process described above captures how users’ urgency increases when they are not assigned a trip and resets to the lowest level once they are served, representing their subsequent trip request.

\begin{figure}[h]
    \centering
    \includegraphics[width=0.86\linewidth]{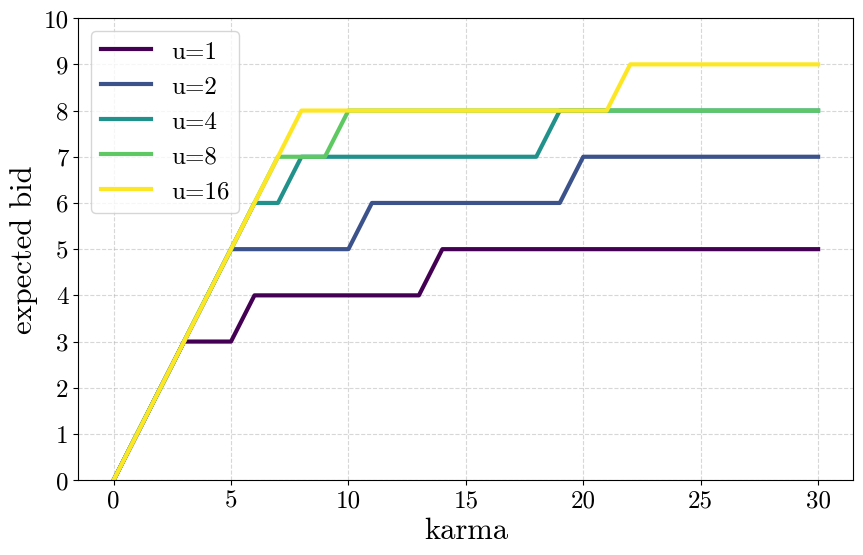}
    \caption{Stationary Nash Equilibrium policy with the urgency transition process in (\ref{eq:phi1})}
    \label{fig:policy1}
\end{figure}

\subsection{Results and discussion}
Figure \ref{fig:policy1} illustrates the Stationary Nash Equilibrium policy played by the users in the proposed augmented Karma-game. As the figure reveals, there is a clear and intuitive trend: as the urgency level of the agents increases, the corresponding bid values also parsimoniously increase. This behavior is consistent with the agents' strategic incentives to secure better outcomes in more critical situations, without bidding too many karma points if it is not necessary. The monotonic relationship between urgency and bid magnitude reflects the rational nature of the equilibrium strategy, where agents balance the immediate need for resource allocation against the long-term value of preserving their karma balances for future high-urgency scenarios.

\begin{figure}[tb]
    \centering
    \includegraphics[width=0.9\linewidth]{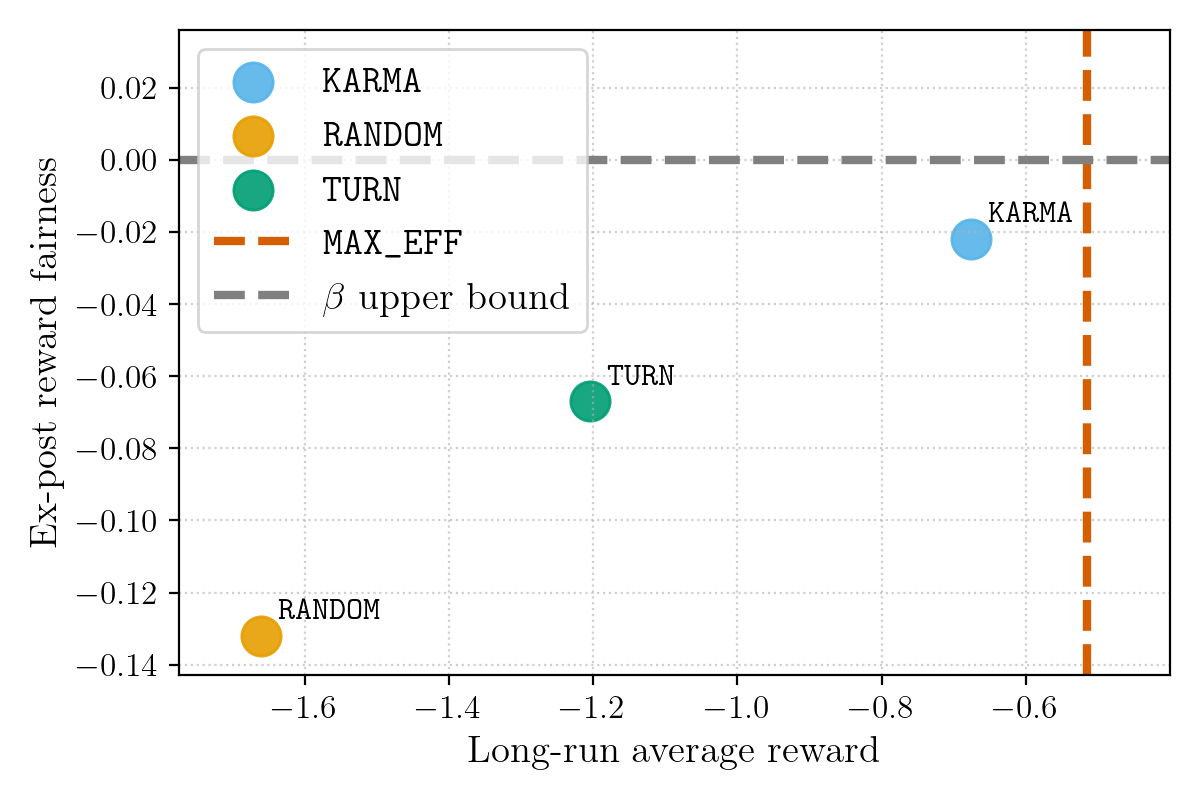}
    \caption{Performance of the augmented Karma economy and the three benchmarks \texttt{RANDOM}, \texttt{TURN}, and \texttt{MAX\_EFF} across the performance metrics introduced in Section~\ref{subsec:bench}.}
    \label{fig:comparison}
\end{figure}

Next, we proceed to evaluate the performance of our proposed approach by comparing the long-run average reward and the ex-post reward fairness of our method with the three benchmark methods introduced in Section~\ref{subsec:bench}. As shown in Figure \ref{fig:comparison}, the Karma-based approach introduced in this work significantly outperforms the \texttt{RANDOM} and \texttt{TURN} policies, both in terms of long-run average reward in the system and ex-post reward fairness. This demonstrates the effectiveness of the Karma mechanism in guiding agents toward more efficient and socially beneficial bidding strategies. Moreover, the long-run average reward achieved by our method is nearly on par with \texttt{MAX\_EFF}, which assumes full knowledge of the system and the ability to centrally coordinate all agents. The fact that our decentralized approach achieves results so close to the upper bound highlights the robustness and efficiency of our design, even in the absence of centralized control.

%% file: 5-conclusions.tex
In this paper, we have presented a novel extension of the Karma economy framework \mods{as a first step toward its application to} Mobility-on-Demand systems, particularly ride-hailing platforms, by incorporating endogenous urgency processes for users. Our work models user time-sensitivity as a dynamic process that evolves in response to system conditions, such as traffic, demand fluctuations, and service availability. This enhancement allows for a more realistic representation of \mods{some aspects of} user behavior in transportation systems, which is essential for designing fair and efficient mobility 
solutions.
Through the development of a game-theoretic model, we have shown how the augmented Karma framework can maintain the theoretical guarantees of efficiency and fairness, while accommodating the dynamic and endogenous nature of user urgency.
\mods{The proof-of-concept numerical simulations in simplified ride-hailing scenarios suggest promising behavior of the framework in terms of system efficiency and} fair access to mobility services.

\mods{Looking ahead, an important direction for future research is to progressively increase the realism of the simulation model, by incorporating the geographical dimension of vehicle rebalancing and relaxing the stationarity assumption, which may be limiting in realistic MoD settings where demand patterns and user urgency vary rapidly throughout the day. Further extensions include the incorporation of heterogeneous user preferences and behavioral biases into the model. Finally, integrating real-world data from existing ride-hailing platforms would allow for a comprehensive validation of the approach in practical settings.}